\pdfoutput=1

\documentclass[12pt,a4paper]{article}

\usepackage{ifthen} 
\newboolean{pdflatex}
\setboolean{pdflatex}{true} 

\newboolean{prl}
\setboolean{prl}{false}

\newboolean{articletitles}
\setboolean{articletitles}{true} 

\newboolean{uprightparticles}
\setboolean{uprightparticles}{false} 

\newboolean{inbibliography}
\setboolean{inbibliography}{false} 

\usepackage[top=1in, bottom=1.25in, left=1in, right=1in]{geometry}

\usepackage{microtype}
\usepackage{lineno}  
\usepackage{xspace} 
\usepackage{caption} 

\usepackage{graphicx}  
\usepackage{color}
\usepackage{colortbl}
\graphicspath{{./figs/}} 

\usepackage{amsmath} 
\usepackage{amssymb}
\usepackage{amsfonts}
\usepackage{upgreek} 

\newcommand*\patchAmsMathEnvironmentForLineno[1]{%
\expandafter\let\csname old#1\expandafter\endcsname\csname #1\endcsname
\expandafter\let\csname oldend#1\expandafter\endcsname\csname
end#1\endcsname
 \renewenvironment{#1}%
   {\linenomath\csname old#1\endcsname}%
   {\csname oldend#1\endcsname\endlinenomath}%
}
\newcommand*\patchBothAmsMathEnvironmentsForLineno[1]{%
  \patchAmsMathEnvironmentForLineno{#1}%
  \patchAmsMathEnvironmentForLineno{#1*}%
}
\AtBeginDocument{%
\patchBothAmsMathEnvironmentsForLineno{equation}%
\patchBothAmsMathEnvironmentsForLineno{align}%
\patchBothAmsMathEnvironmentsForLineno{flalign}%
\patchBothAmsMathEnvironmentsForLineno{alignat}%
\patchBothAmsMathEnvironmentsForLineno{gather}%
\patchBothAmsMathEnvironmentsForLineno{multline}%
\patchBothAmsMathEnvironmentsForLineno{eqnarray}%
}

\usepackage{hyperref}    
\usepackage[all]{hypcap} 

\usepackage{xspace} 
\usepackage{upgreek}

\def\lhcb {\mbox{LHCb}\xspace}

\def\lhc    {\mbox{LHC}\xspace}

\def\MagUp {\mbox{\em Mag\kern -0.05em Up}\xspace}
\def\MagDown {\mbox{\em MagDown}\xspace}

\ifthenelse{\boolean{uprightparticles}}%
{

 \def\Ppi         {\ensuremath{\uppi}\xspace}

 \def\PDelta      {\ensuremath{\Delta}\xspace}                 
 \def\PXi      {\ensuremath{\Xi}\xspace}                 
 \def\PLambda      {\ensuremath{\Lambda}\xspace}                 
 \def\PSigma      {\ensuremath{\Sigma}\xspace}                 
 \def\POmega      {\ensuremath{\Omega}\xspace}                 
 \def\PUpsilon      {\ensuremath{\Upsilon}\xspace}                 
 
 \def\PB      {\ensuremath{\mathrm{B}}\xspace}                 
                  
 \def\PD      {\ensuremath{\mathrm{D}}\xspace}

 \def\PK      {\ensuremath{\mathrm{K}}\xspace}

 \def\Pb      {\ensuremath{\mathrm{b}}\xspace}                 
 \def\Pc      {\ensuremath{\mathrm{c}}\xspace}

 \def\Pi      {\ensuremath{\mathrm{i}}\xspace}

 \def\Ps      {\ensuremath{\mathrm{s}}\xspace}

}
{

 \def\Ppi         {\ensuremath{\pi}\xspace}

 \mathchardef\PDelta="7101
 \mathchardef\PXi="7104
 \mathchardef\PLambda="7103
 \mathchardef\PSigma="7106
 \mathchardef\POmega="710A
 \mathchardef\PUpsilon="7107
                  
 \def\PB      {\ensuremath{B}\xspace}                 
                  
 \def\PD      {\ensuremath{D}\xspace}

 \def\PK      {\ensuremath{K}\xspace}

 \def\Pb      {\ensuremath{b}\xspace}                 
 \def\Pc      {\ensuremath{c}\xspace}

 \def\Pi      {\ensuremath{i}\xspace}

 \def\Ps      {\ensuremath{s}\xspace}

}

\makeatletter
\ifcase \@ptsize \relax
  \newcommand{\miniscule}{\@setfontsize\miniscule{4}{5}}
\or
  \newcommand{\miniscule}{\@setfontsize\miniscule{5}{6}}
\or
  \newcommand{\miniscule}{\@setfontsize\miniscule{5}{6}}
\fi
\makeatother

\DeclareRobustCommand{\optbar}[1]{\shortstack{{\miniscule (\rule[.5ex]{1.25em}{.18mm})}
  \\ [-.7ex] $#1$}}

\def\squark    {{\ensuremath{\Ps}}\xspace}

\def\cquark    {{\ensuremath{\Pc}}\xspace}

\def\bquark    {{\ensuremath{\Pb}}\xspace}

\def\pion   {{\ensuremath{\Ppi}}\xspace}

\def\pip    {{\ensuremath{\pion^+}}\xspace}
\def\pim    {{\ensuremath{\pion^-}}\xspace}

\def\kaon    {{\ensuremath{\PK}}\xspace}
  \def\Kbar    {{\kern 0.2em\overline{\kern -0.2em \PK}{}}\xspace}

\def\KorKbar    {\kern 0.18em\optbar{\kern -0.18em K}{}\xspace}

\def\Kp      {{\ensuremath{\kaon^+}}\xspace}
\def\Km      {{\ensuremath{\kaon^-}}\xspace}

  \def\Dbar    {{\kern 0.2em\overline{\kern -0.2em \PD}{}}\xspace}
\def\D       {{\ensuremath{\PD}}\xspace}

\def\DorDbar    {\kern 0.18em\optbar{\kern -0.18em D}{}\xspace}
\def\Dz      {{\ensuremath{\D^0}}\xspace}
\def\Dzb     {{\ensuremath{\Dbar{}^0}}\xspace}

\def\Dstarp  {{\ensuremath{\D^{*+}}}\xspace}
\def\Dstarm  {{\ensuremath{\D^{*-}}}\xspace}
\def\Dstarpm {{\ensuremath{\D^{*\pm}}}\xspace}

\def\Bbar    {{\ensuremath{\kern 0.18em\overline{\kern -0.18em \PB}{}}}\xspace}

\def\BorBbar    {\kern 0.18em\optbar{\kern -0.18em B}{}\xspace}

  \def\Y#1S{\ensuremath{\PUpsilon{(#1S)}}\xspace}

\def\Lbar        {{\ensuremath{\kern 0.1em\overline{\kern -0.1em\PLambda}}}\xspace}
\def\LorLbar    {\kern 0.18em\optbar{\kern -0.18em \PLambda}{}\xspace}

\newcommand{\decay}[2]{\ensuremath{#1\!\to #2}\xspace}         

\def\to                 {\ensuremath{\rightarrow}\xspace}

\def\CP                {{\ensuremath{C\!P}}\xspace}

\def\AT#1     {\ensuremath{A_{\mathrm{T}}^{#1}}\xspace}           

\def\C#1      {\ensuremath{\mathcal{C}_{#1}}\xspace}                       
\def\Cp#1     {\ensuremath{\mathcal{C}_{#1}^{'}}\xspace}                    
\def\Ceff#1   {\ensuremath{\mathcal{C}_{#1}^{\mathrm{(eff)}}}\xspace}        
\def\Cpeff#1  {\ensuremath{\mathcal{C}_{#1}^{'\mathrm{(eff)}}}\xspace}       
\def\Ope#1    {\ensuremath{\mathcal{O}_{#1}}\xspace}                       
\def\Opep#1   {\ensuremath{\mathcal{O}_{#1}^{'}}\xspace}                    

\def\agamma     {\ensuremath{A_{\Gamma}}\xspace}

\newcommand{\ket}[1]{\ensuremath{|#1\rangle}}              

\newcommand{\tev}{\ifthenelse{\boolean{inbibliography}}{\ensuremath{~T\kern -0.05em eV}\xspace}{\ensuremath{\mathrm{\,Te\kern -0.1em V}}}\xspace}
\newcommand{\gev}{\ensuremath{\mathrm{\,Ge\kern -0.1em V}}\xspace}
\newcommand{\mev}{\ensuremath{\mathrm{\,Me\kern -0.1em V}}\xspace}
\newcommand{\kev}{\ensuremath{\mathrm{\,ke\kern -0.1em V}}\xspace}
\newcommand{\ev}{\ensuremath{\mathrm{\,e\kern -0.1em V}}\xspace}
\newcommand{\gevc}{\ensuremath{{\mathrm{\,Ge\kern -0.1em V\!/}c}}\xspace}
\newcommand{\mevc}{\ensuremath{{\mathrm{\,Me\kern -0.1em V\!/}c}}\xspace}
\newcommand{\gevcc}{\ensuremath{{\mathrm{\,Ge\kern -0.1em V\!/}c^2}}\xspace}
\newcommand{\gevgevcccc}{\ensuremath{{\mathrm{\,Ge\kern -0.1em V^2\!/}c^4}}\xspace}
\newcommand{\mevcc}{\ensuremath{{\mathrm{\,Me\kern -0.1em V\!/}c^2}}\xspace}

\def\invfb   {\ensuremath{\mbox{\,fb}^{-1}}\xspace}

\def\ps   {\ensuremath{{\mathrm{ \,ps}}}\xspace}

\newcommand{\chisq}{\ensuremath{\chi^2}\xspace}

\newcommand{\chisqip}{\ensuremath{\chi^2_{\text{IP}}}\xspace}

\def\deriv {\ensuremath{\mathrm{d}}}

\def\gsim{{~\raise.15em\hbox{$>$}\kern-.85em
          \lower.35em\hbox{$\sim$}~}\xspace}
\def\lsim{{~\raise.15em\hbox{$<$}\kern-.85em
          \lower.35em\hbox{$\sim$}~}\xspace}

\def\tell1  {TELL1\xspace}
\def\ukl1   {UKL1\xspace}

\newcommand{\ie}{\mbox{\itshape i.e.}\xspace}

\usepackage{cite} 
\usepackage{mciteplus}

\newcommand{\aprompt}{\ensuremath{A_{\rm prompt}(t)}\xspace}
\newcommand{\asec}{\ensuremath{A_{\rm sec}(t)}\xspace}
\newcommand{\fsec}{\ensuremath{f_{\rm sec}(t)}\xspace}
\usepackage{longtable} 

\usepackage{tabularx}
\usepackage{multirow}
\usepackage{epstopdf}

\DeclareGraphicsExtensions{.pdf,.PDF,.png,.PNG}

\begin{document}

\renewcommand{\thefootnote}{\fnsymbol{footnote}}
\setcounter{footnote}{1}


\begin{titlepage}
\pagenumbering{roman}

\vspace*{-1.5cm}
\centerline{\large EUROPEAN ORGANIZATION FOR NUCLEAR RESEARCH (CERN)}
\vspace*{1.5cm}
\noindent
\begin{tabular*}{\linewidth}{lc@{\extracolsep{\fill}}r@{\extracolsep{0pt}}}
\ifthenelse{\boolean{pdflatex}}
{\vspace*{-2.7cm}\mbox{\!\!\!\includegraphics[width=.14\textwidth]{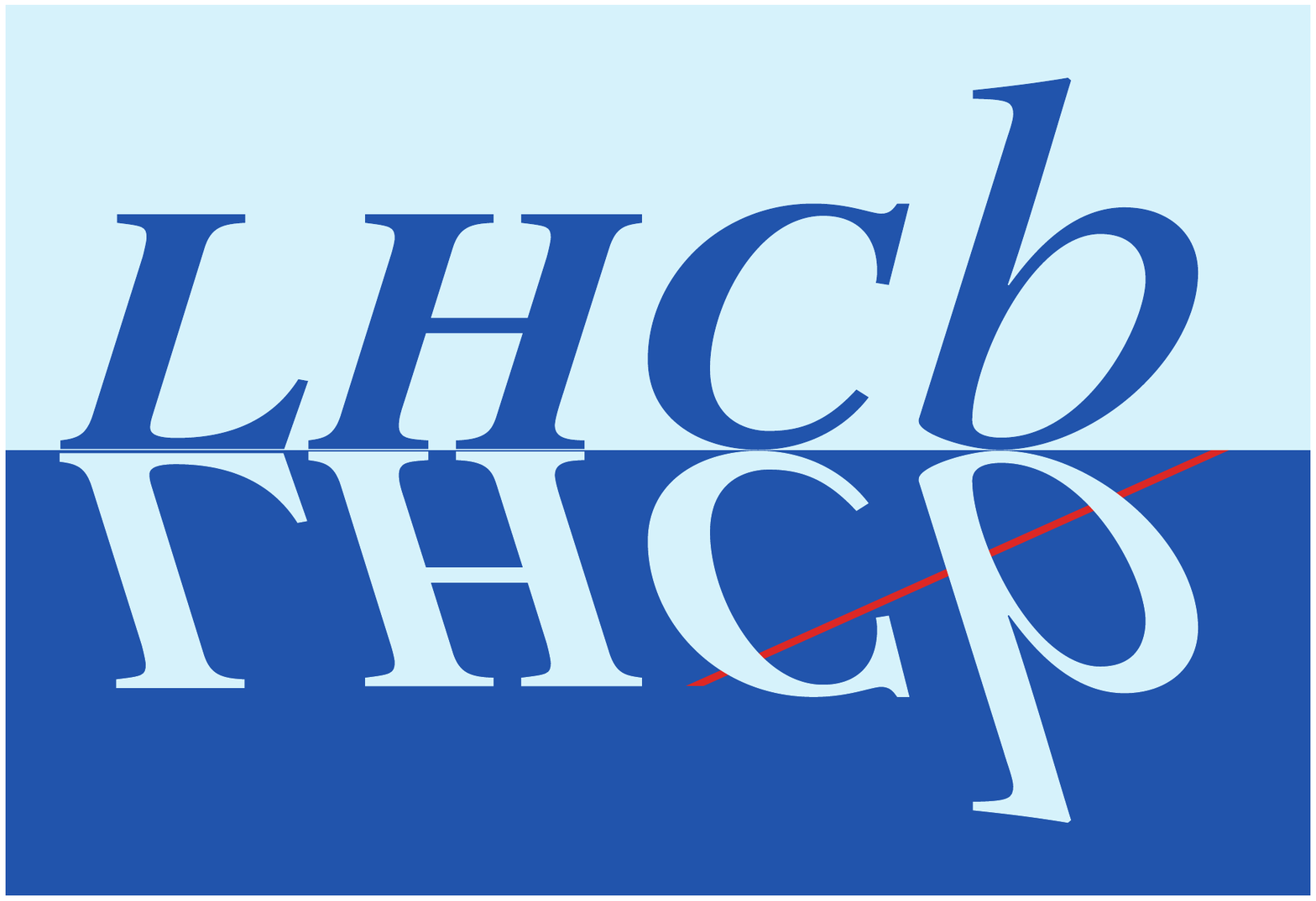}} & &}%
{\vspace*{-1.2cm}\mbox{\!\!\!\includegraphics[width=.12\textwidth]{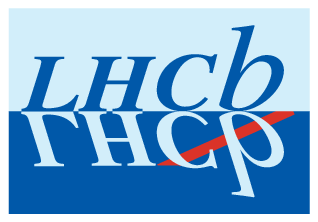}} & &}%
\\
 & &  CERN-EP-2017-028 \\  
 & & LHCb-PAPER-2016-063 \\  
 & & August 7, 2017 \\ 
\end{tabular*}

\vspace*{2.0cm}

{\normalfont\bfseries\boldmath\huge
\begin{center}
Measurement of the \CP violation parameter \agamma in \decay{\Dz}{\Kp \Km}  and  \decay{\Dz}{\pip \pim} decays
\end{center}
}

\vspace*{1.cm}

\begin{center}
The LHCb collaboration\footnote{Authors are listed at the end of this Letter.}
\end{center}


\begin{abstract}
\noindent
Asymmetries in the time-dependent rates of \decay{\Dz}{\Kp \Km} and \decay{\Dz}{\pip \pim} 
decays are measured in a $pp$ collision data sample collected with the \lhcb detector during \lhc Run~1, corresponding to an integrated luminosity of $3\invfb$.
The asymmetries in effective decay widths between 
\Dz and \Dzb decays, sensitive to indirect \CP violation, are
measured to be
$\agamma(\Kp \Km) = (-0.30 \pm 0.32 \pm 0.10)\times 10^{-3}$ and 
$\agamma(\pip \pim) = (0.46 \pm 0.58 \pm 0.12)\times 10^{-3}$, 
where the first uncertainty is statistical and the second systematic.
These measurements show no evidence for \CP violation and improve
on the precision of the previous best measurements by nearly a factor of two.

\end{abstract}

\vspace*{1.0cm}

\begin{center}
  Published in Phys.~Rev.~Lett. 118, 261803 (2017)
\end{center}

\vspace{\fill}

{\footnotesize 
\centerline{\copyright~CERN on behalf of the \lhcb collaboration, licence \href{http://creativecommons.org/licenses/by/4.0/}{CC-BY-4.0}.}}
\vspace*{2mm}

\end{titlepage}


\newpage
\setcounter{page}{2}
\mbox{~}

%
%

\cleardoublepage


\renewcommand{\thefootnote}{\arabic{footnote}}
\setcounter{footnote}{0}



\pagestyle{plain} 
\setcounter{page}{1}
\pagenumbering{arabic}


%


Symmetry under the combined operations of charge conjugation and parity (\CP) was found to be violated in flavor-changing interactions of the \squark quark~\cite{Christenson:1964fg}, and later in processes
involving the \bquark quark~\cite{Aubert:2001nu,Abe:2001xe}. 
Within the Standard Model, violation of \CP symmetry in the charm 
sector is predicted at a level below $\mathcal{O}(10^{-3})$~\cite{Bobrowski:2010xg,Grossman:2006jg}.
Charm hadrons are the only particles where \CP violation  involving up-type quarks is  expected to be observable, 
providing a unique opportunity to detect  effects beyond the Standard Model that leave down-type quarks unaffected.

A sensitive probe of \CP violation in the charm sector is given by decays of \Dz mesons
into \CP eigenstates $f$, where $f=\pi^+\pi^-$ or $f=K^+K^-$.
The time-integrated \CP asymmetries and
the charm mixing parameters $x \equiv (m_2 -m_1)/\Gamma$ and $y \equiv (\Gamma_2 - \Gamma_1 )/(2\Gamma)$~\cite{HFAG},
where $m_{1,2}$ and $\Gamma_{1,2}$ are the masses and widths of the mass eigenstates \ket{\D_{1,2}},
are known to be small~\cite{LHCb-PAPER-2013-053,Aaltonen:2011se,LHCb-PAPER-2016-035}.
As a result, the time-dependent \CP asymmetry of each decay mode can be approximated as~\cite{Aaltonen:2011se}
\begin{equation}
A_{\CP}(t) \equiv \frac{\Gamma(\decay{\Dz(t)}{ f}) - %
 	 			\Gamma(\decay{\Dzb(t)}{ f})}%
				{\Gamma(\decay{\Dz(t)}{ f}) + %
 				\Gamma(\decay{\Dzb(t)}{ f})} 
		\simeq a^f_\textup{dir} -\agamma \frac{t}{\tau_D}, 
\label{eq:Aind}
\end{equation}
where $\Gamma (\Dz(t) \to f)$ and $\Gamma (\Dzb(t) \to f)$ indicate
the time-dependent decay rates of an initial \Dz or \Dzb decaying to a final state $f$ at decay time $t$,
$\tau_D = 1/\Gamma = 2/(\Gamma_1 + \Gamma_2)$ is the average lifetime of the \Dz meson, 
$a^{f}_\textup{dir}$ is the asymmetry related to direct \CP violation and
$\agamma$ is the asymmetry between the \Dz and \Dzb effective decay widths,
\begin{equation}\label{eq:A_G}
\agamma \equiv \frac{\hat{\Gamma}_{\decay{\Dz}{ f}} - %
 	 			\hat{\Gamma}_{\decay{\Dzb}{ f}}}%
				{\hat{\Gamma}_{\decay{\Dz}{ f}} + %
 				\hat{\Gamma}_{\decay{\Dzb}{ f}}}. 
\end{equation}
The effective decay width $\hat{\Gamma}_{\decay{\Dz}{ f}}$ 
is defined as $  \int_0^\infty \Gamma( \Dz(t) \to f) \, \deriv t  /
\int_0^\infty t \, \Gamma(\Dz(t) \to f)\, \deriv t$, \ie the inverse of
the effective lifetime.

Neglecting contributions from subleading 
amplitudes~\cite{Grossman:2006jg, Du:2006jc},
$a^{f}_\textup{dir}$ vanishes and \agamma is independent of the final state $f$.
Furthermore, in the absence of \CP violation in mixing, it can be found that 
$\agamma = - x \sin \phi $, 
where  $\phi = \arg{((q\overline{A}_f)/(pA_f))}$,
$A_f(\overline{A}_f)$ is the amplitude of the \decay{\Dz}{f}(\decay{\Dzb}{f}) decay,
and $p$ and $q$ are the coefficients of the decomposition of the mass eigenstates $\ket{\D_{1,2}} = p \ket{\Dz} \pm q \ket{\Dzb}$.
This implies that $|\agamma| <  |x| \lsim 5\times 10^{-3}$~\cite{HFAG}.

This Letter presents a measurement of \agamma with $pp$ collision data
collected by \lhcb in Run~1, corresponding to an integrated luminosity of $3\invfb$,
with $1\invfb$ collected during 2011 at a center-of-mass energy of
$7\tev$ and $2\invfb$ collected during 2012 at $8\tev$. 
The measurements presented are independent of the center-of-mass
energy, but the two periods are analyzed separately  to account for
differences in cross-sections and in the general running conditions.
The charge of the pion from the 
$\decay{\Dstarp}{ \Dz \pip}$ ($\decay{\Dstarm}{ \Dzb \pim}$) decay is used to identify the 
flavor of the \Dz\ (\Dzb) meson at production.
Two different approaches are used to perform the measurement of \agamma.
The first is a new method based on Eq.~\eqref{eq:Aind}  and provides
the more precise results. This is described in the following text, 
unless otherwise stated. 
The other method,  based on Eq.~\eqref{eq:A_G}, has been described
previously in Ref.~\cite{LHCb-PAPER-2013-054} and is only summarized here.
In the following, inclusion of charge-conjugate processes is implied throughout, unless otherwise stated.
%

The \lhcb detector~\cite{Alves:2008zz,LHCb-DP-2014-002} is a
single-arm forward spectrometer covering the \mbox{pseudorapidity} 
range $2<\eta <5$, designed for the study of particles containing
\bquark or \cquark quarks.
The detector includes a high-precision tracking system
consisting of a silicon-strip vertex detector, surrounding the $pp$ interaction region 
and allowing  \cquark hadrons to be identified 
by their characteristic 
flight distance, a large-area silicon-strip detector located
upstream of a dipole magnet with a bending power of about $4\,\rm{Tm}$, and three stations of silicon-strip detectors and straw drift tubes 
placed downstream of the magnet.
Two ring-imaging Cherenkov detectors 
provide particle identification to distinguish kaons from pions. 
The polarity of the dipole magnet is periodically reversed during data taking. 
The configuration with the magnetic field vertically upwards (downwards), \MagUp (\MagDown), 
bends positively (negatively) charged particles in the horizontal plane towards the center of 
the Large Hadron Collider. 
The \lhcb coordinate system is a right-handed system, with the $z$ axis pointing along the beam direction, 
$y$ pointing vertically upwards, and $x$ pointing in the horizontal direction
away from the collider center.



An online event selection is performed by a trigger system~\cite{LHCb-DP-2012-004}, 
consisting of a hardware stage, based on information from the calorimeter and muon
systems, followed by a software stage, which applies a full event reconstruction.
All events passing the hardware trigger are analysed. 
Both the software trigger and the subsequent event selection use
kinematic and  topological variables to separate signal decays from background.
In the software trigger, two oppositely charged particles are required to form a \Dz candidate that is significantly displaced from
any primary $pp$ interaction vertex (PV) in the event, 
and at least one of these two particles must have a minimum momentum transverse to the beam direction of $1.7\gevc$ or $1.6\gevc$ depending 
on the running conditions.
The \Dz candidates are combined with all
possible pion candidates (``soft pions'') to form \Dstarp candidates.
No requirements are imposed on the soft pions at trigger level.

Offline requirements are placed on:  
the \Dstarp vertex fit quality, where the vertex 
formed by the \Dz and the soft $\pip$ candidate is constrained to coincide with a PV; the \Dz flight distance and transverse momentum; 
the angle between the \Dz  momentum and the vector from the PV to the \Dz decay vertex;
the $\chisqip$ value of each of the \Dz  decay products,
where $\chisqip$  is defined as the difference between the vertex fit $\chi^2$ 
of a PV reconstructed with and without the particle under consideration. 
The two signal samples, $\pip\pim$ and $\Kp\Km$, plus the 
Cabibbo-favored $\Km\pip$ control
sample, are defined imposing further requirements on the particle identification likelihood, which 
is calculated from a combination of information from the Cherenkov detectors and the tracking system~\cite{LHCb-DP-2012-003}.
About 13\% of the selected events have more than one candidate, 
mostly due to a single \Dz candidate being associated with
multiple soft pions. One of those candidates is then selected at random.

The \Dz\ signal region is defined by the requirement that the invariant mass be within $\pm 24 \mevcc$  (approximately $\pm3$ times the mass resolution)
of the known value~\cite{PDG2016}. 
The reconstructed decay times of charm mesons that originate 
from weak decays of \bquark hadrons (secondary decays) are biased towards positive values, 
and thus these decays are treated as background.
This contamination is reduced to a few percent by requiring the reconstructed \Dz momentum to point back to the PV and $\chisqip(\Dz)<9$.
A systematic uncertainty on the final measurement is assigned due to residual secondary background.
The signal yields of the \Kp\Km, \pip\pim  and \Km\pip  samples, obtained by fitting 
the distributions of the invariant mass difference $\Delta m \equiv m(\Dz\pip) - m(\Dz)$,  
are reported in Table~\ref{tab:final_yields}.
A Johnson $S_U$-distribution~\cite{Johnson:1949zj} 
plus the sum of three Gaussian functions is used to model the signal, while 
the background is described by an empirical function of the form $1 - \exp [(\Delta m - \Delta m_0)/\alpha] + \beta(\Delta m/\Delta m_0 - 1)$, 
where $\Delta m_0$ is the threshold of the function, and $\alpha$ and $\beta$ describe its shape.
\begin{table}[h]
\centering
\caption{Signal yields in millions after all selection requirements.}
\label{tab:final_yields}
\begin{tabular}{l  c  c  c}
\hline
Subsample &  \decay{\Dz}{\Km \pip} & \decay{\Dz}{\Kp \Km} & \decay{\Dz}{\pip \pim}  \\
\hline
2011 \MagUp       & $10.7$ & $ 1.2$ & $ 0.4$\\
2011 \MagDown  & $15.5$ & $ 1.7$ & $ 0.5$\\
2012  \MagUp     & $30.0$ & $ 3.3$ & $ 1.0$\\
2012 \MagDown  & $31.3$ & $ 3.4$ & $ 1.1$\\
\hline
Total  & $87.5$ & $ 9.6$ & $ 3.0$\\
\hline
\end{tabular}
\end{table}

The effect of a small residual background of fake \Dstarp candidates,
dominated by real \Dz decays associated  with uncorrelated pions, is removed by a sideband-subtraction procedure.  
The signal region is defined as $\Delta m \in [144.45, 146.45]\mevcc$, 
about $\pm5$ times the $\Delta m$ resolution, and the 
sideband region as $\Delta m \in [149, 154]\mevcc$.
The uncertainty associated with this procedure is accounted for  within the systematic uncertainty.

\begin{figure*}[tbh]
\centering
\ifthenelse{\boolean{prl}}
{
\includegraphics[width=0.325\textwidth]{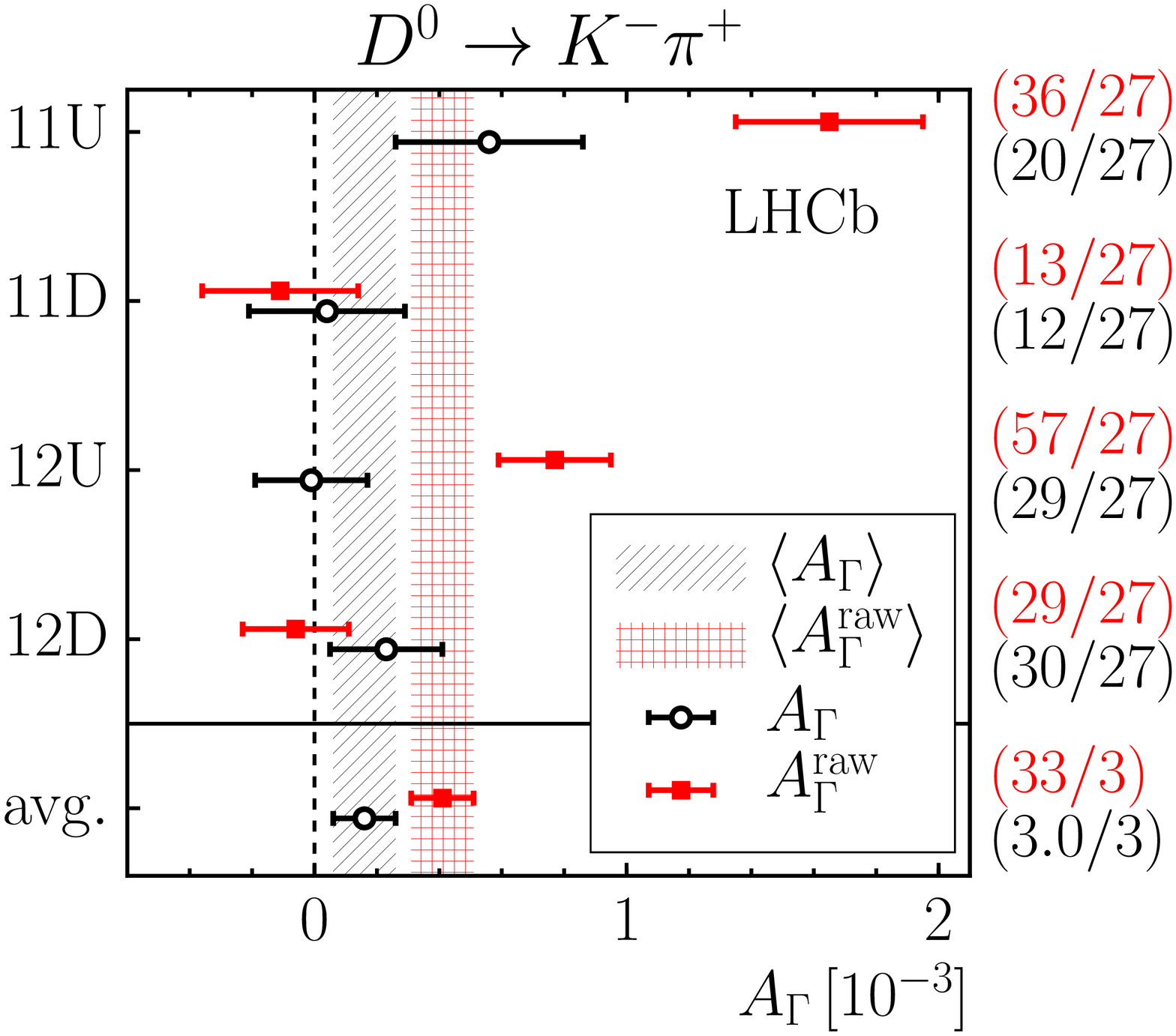}
\includegraphics[width=0.325\textwidth]{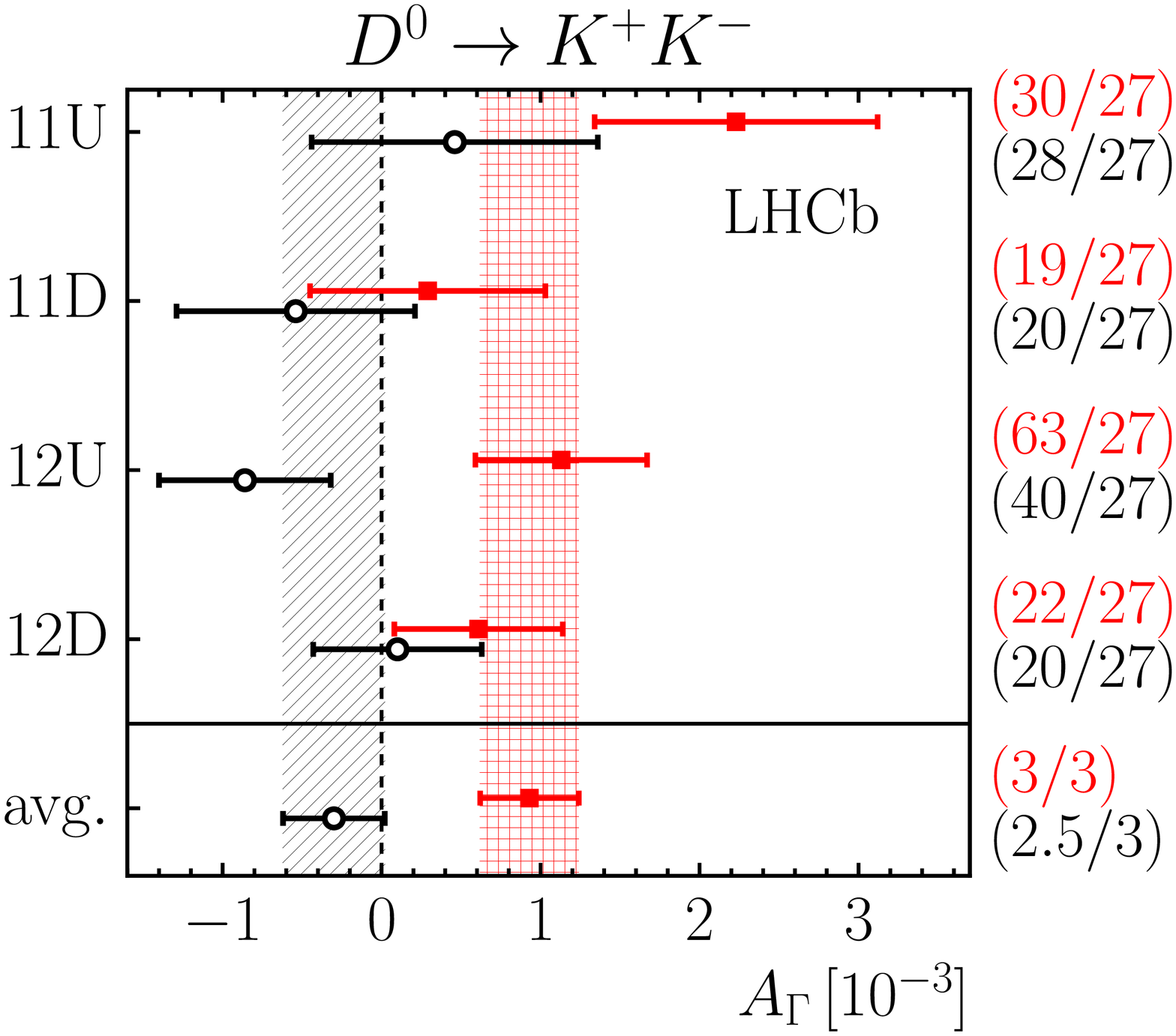}
\includegraphics[width=0.325\textwidth]{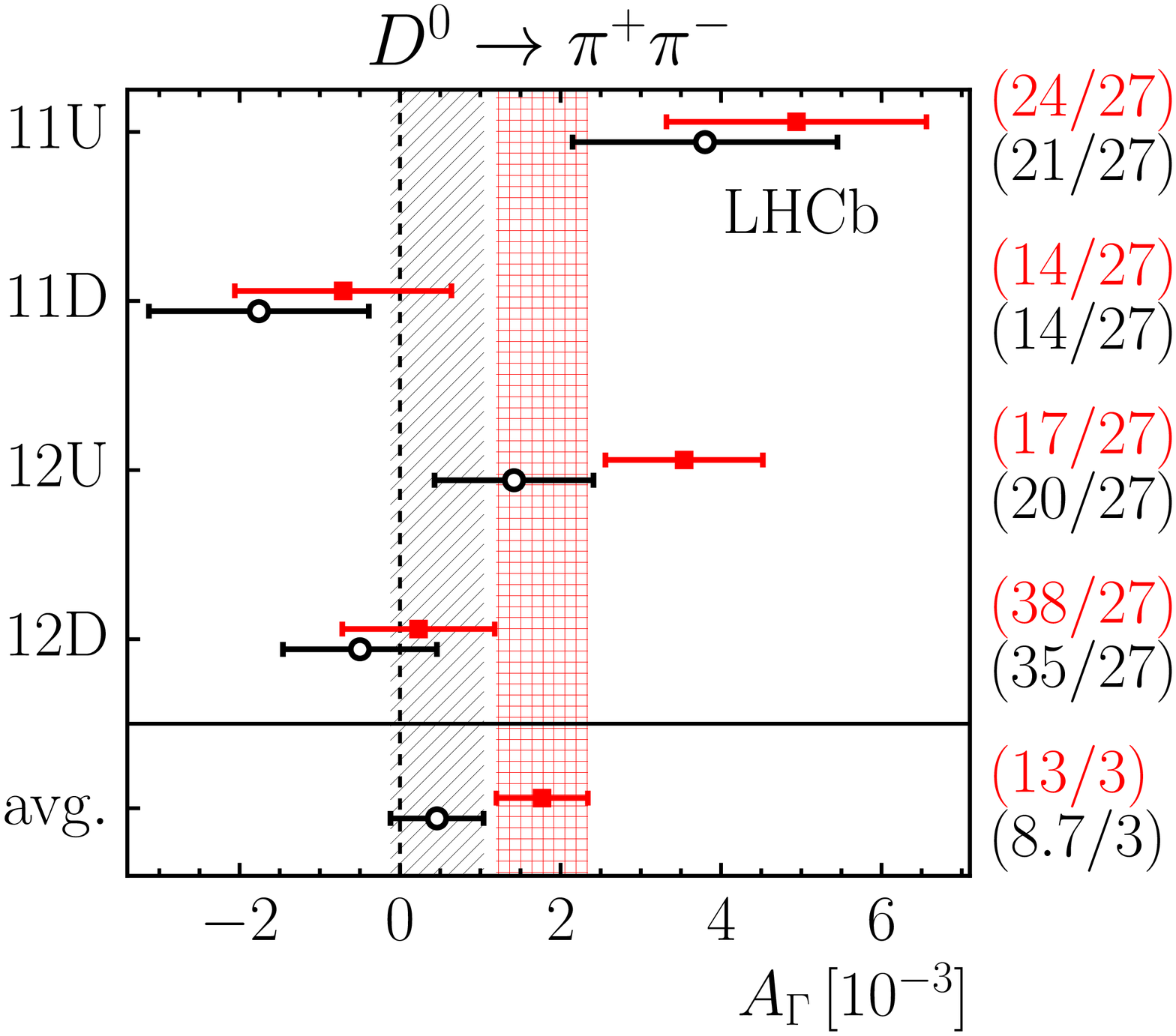}
}
{
\includegraphics[width=0.495\textwidth]{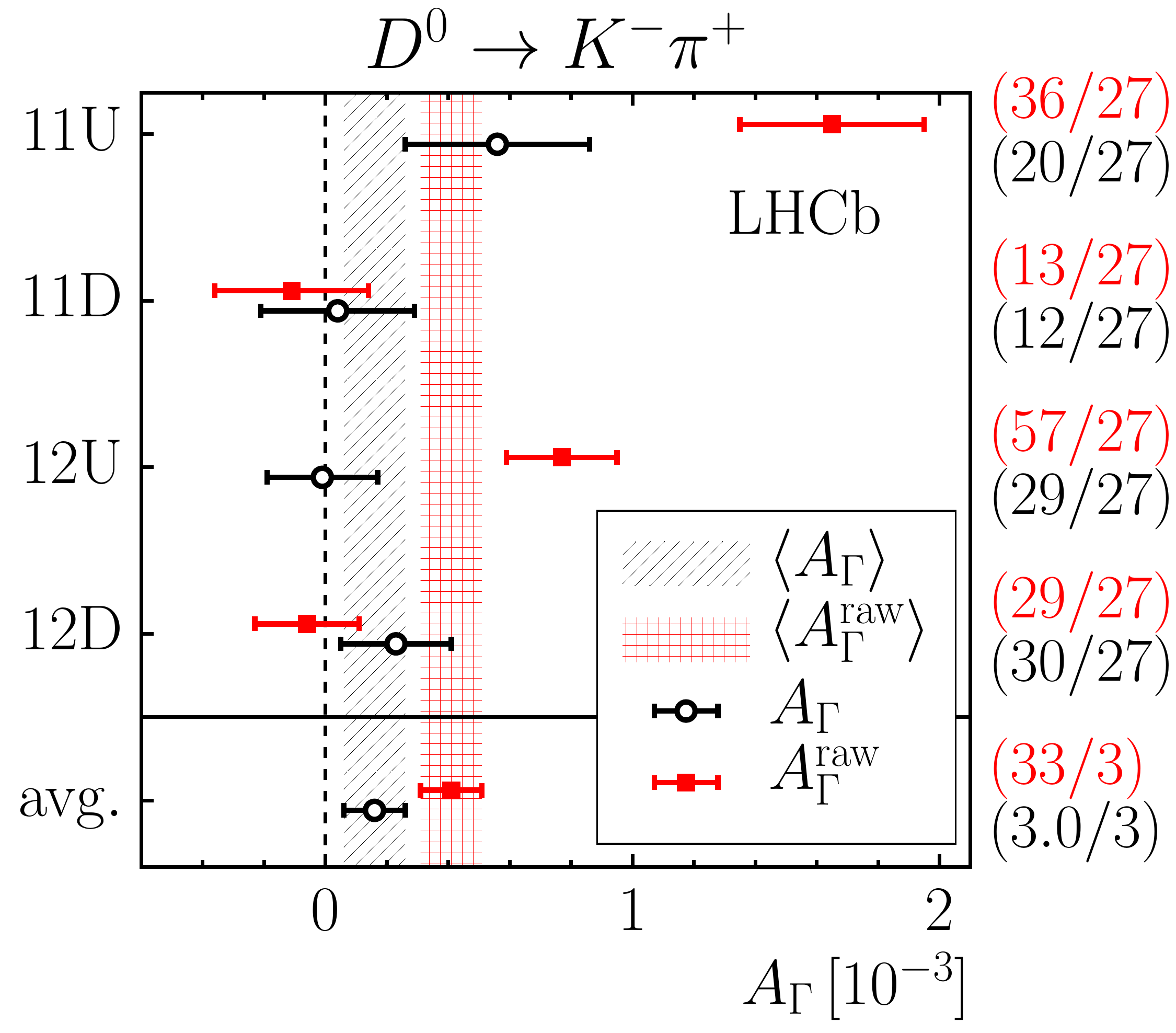} \\
\includegraphics[width=0.495\textwidth]{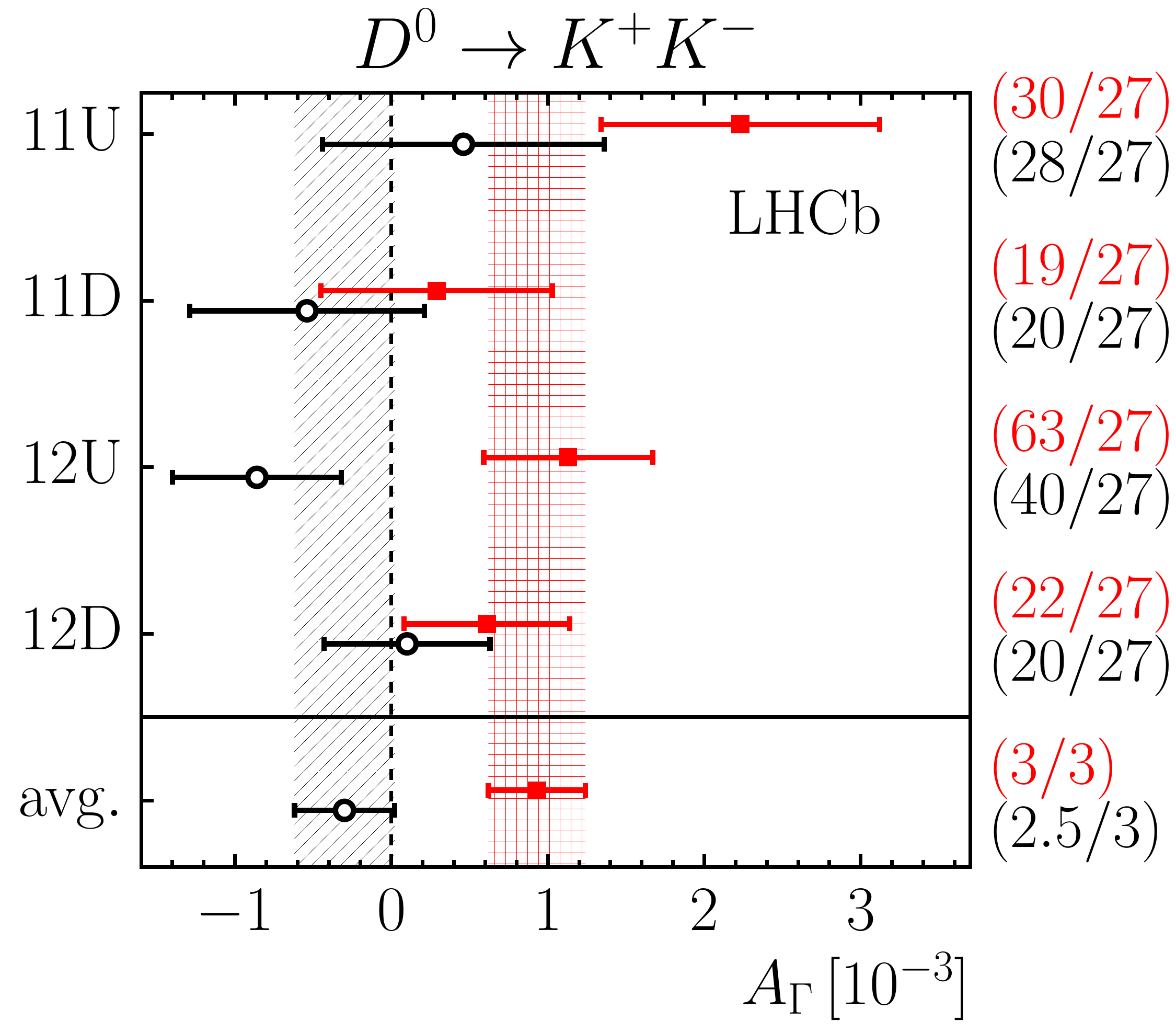}
\includegraphics[width=0.495\textwidth]{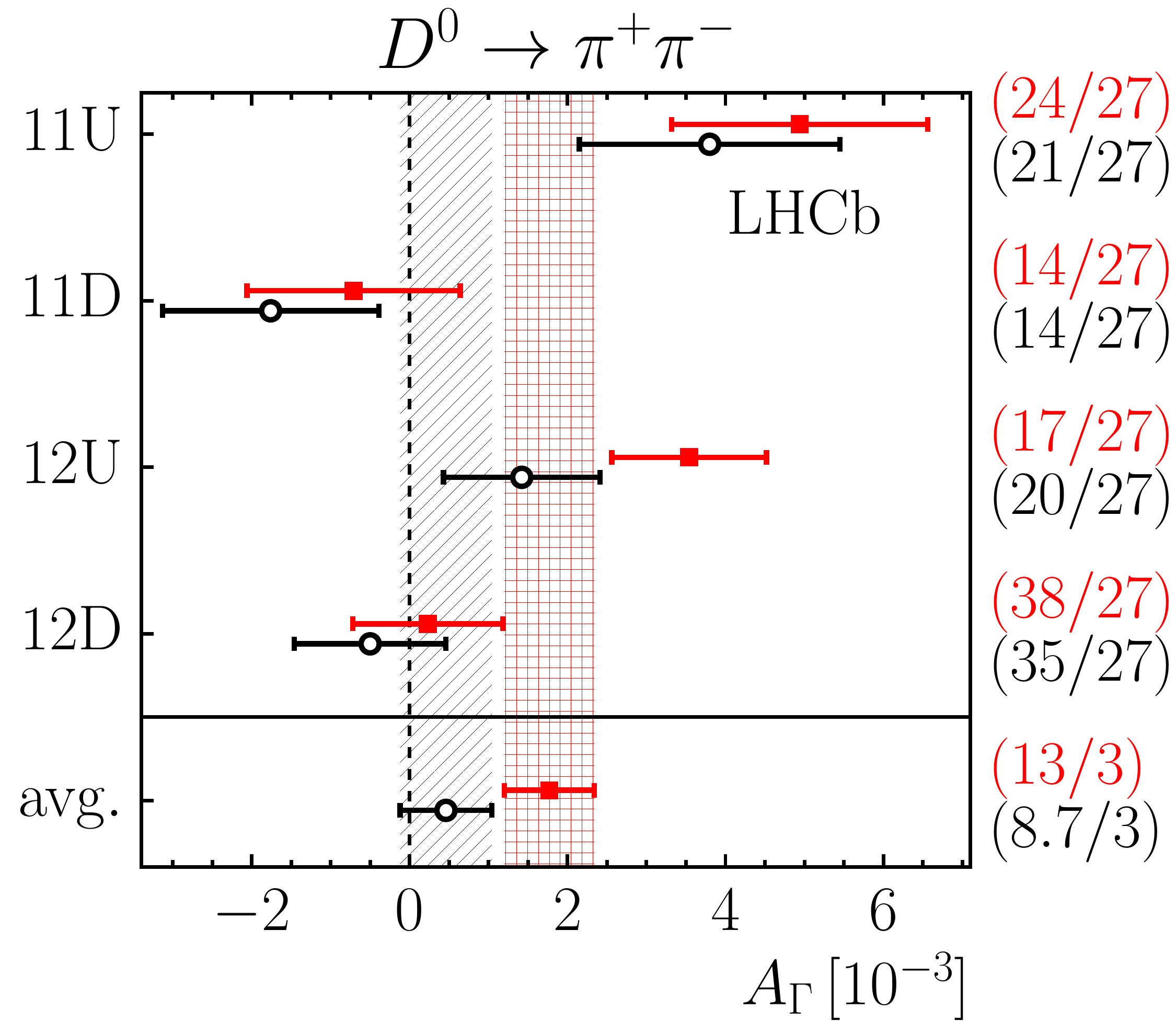}
}
\caption{Results from $\agamma$ fits in each subsample, before (solid red squares) and after (empty black dots) the asymmetry correction.
 Fit qualities ($\chisq/$number of degrees of freedom)  are also reported to the right of each graph. 
The weighted average of the four \agamma values is indicated before (red hatched band) and after (black hatched band) the correction. 
The numerical values for the averages are
$\agamma(\Km\pip) =(0.41 \pm 0.10)\times 10^{-3}$, $\agamma(\Kp\Km) =(0.93 \pm 0.31)\times 10^{-3}$ and 
$\agamma(\pip\pim) =(1.77 \pm 0.57)\times 10^{-3}$ before the correction, and 
$\agamma(\Km\pip) =(0.16 \pm 0.10)\times 10^{-3}$, $\agamma(\Kp\Km) =(-0.30 \pm 0.32)\times 10^{-3}$ and 
$\agamma(\pip\pim) =(0.46 \pm 0.58)\times 10^{-3}$ after the correction.
The label  2011 (2012) is abbreviated 11 (12) and \MagUp (\MagDown) is abbreviated $\mathrm{U}(\mathrm{D})$.}
\label{fig:pseudoAG-raw}
\end{figure*}

The structure of the \lhcb detector is nearly 
symmetric under reflection in the vertical plane containing the beam axis.
Nevertheless, departures from the nominal geometry and variations of
the efficiency in different parts of the detector produce small
residual deviations from an ideally symmetric detector acceptance. 
An important part of the analysis is therefore the determination and correction of 
these residual asymmetries. 
The method to achieve  this is developed by exploiting the large 
control sample available in the $\decay{\Dz}{\Km \pip}$ mode, where the time-dependent asymmetry is expected to be negligible.
The distribution of the \Dz decay time in the range 
$[0.6\tau_\D, 20\tau_\D]$ is divided into $30$ approximately equally populated bins, and the \Dz--\Dzb yield asymmetry after background removal is determined in each of them. {The lower bound is introduced to remove the initial turn-on region of the trigger efficiency, to avoid potential biases due to charge asymmetries of the quickly varying acceptance function}.
The measured asymmetry $A(t)$ is then fitted with a linear function of the decay time in units of $\tau_D$, 
the slope of which is taken as the estimate of $\agamma$ (see Eq.~\eqref{eq:Aind}).
For the \pip\pim and \Kp\Km  final states the slope is kept blind until the completion of the analysis.  
The slope for the \Km\pip sample, expected to be unmeasurably small, 
is not blinded. Figure~\ref{fig:pseudoAG-raw}  shows the values of \agamma  obtained in the  
four subsamples defined in Table~\ref{tab:final_yields}.
The presence of significant deviations from zero for the control channel 
indicates the existence of non-negligible time-dependent residual detector asymmetries. 
They partially cancel in the combination of the \MagUp and \MagDown samples, but not completely, yielding an overall average 
that is incompatible with zero. These residual biases arise due to correlations between the decay time and other 
kinematic variables that affect the efficiency, most notably the momentum of the soft pion. 

\begin{figure}[tbh]
\centering
\includegraphics[width=0.49\columnwidth]{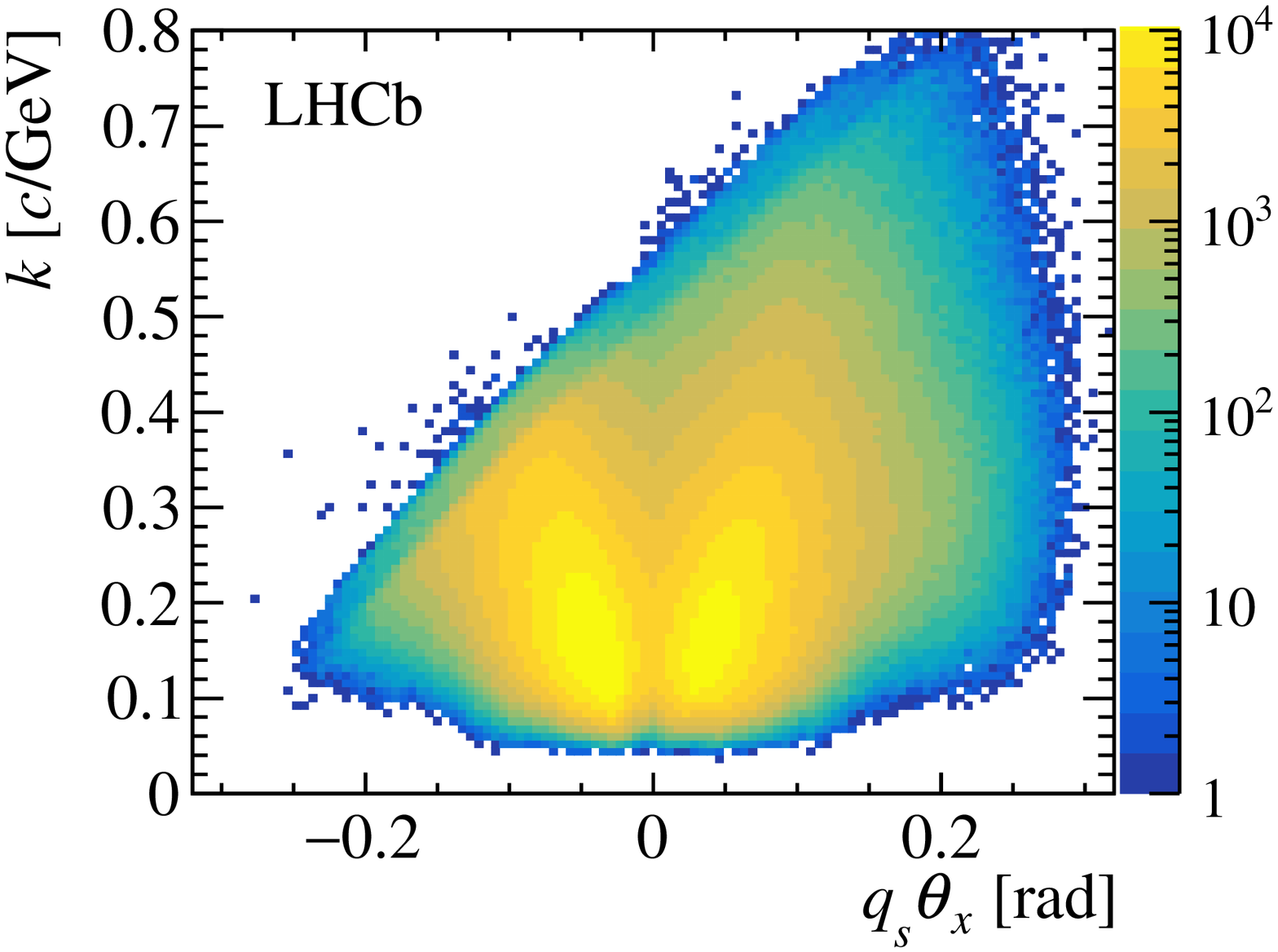}
\includegraphics[width=0.49\columnwidth]{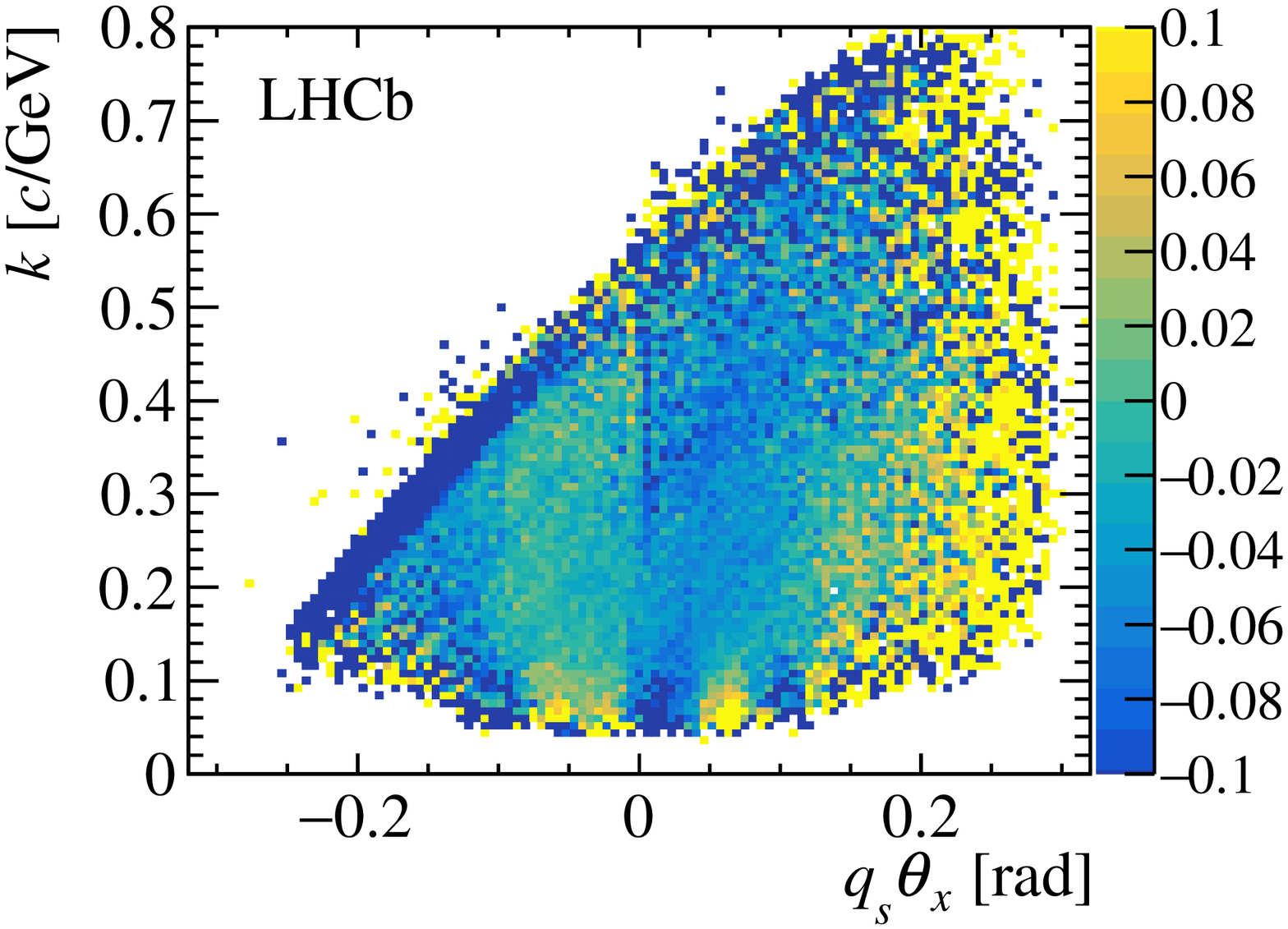}
\caption{(Left) Sum and (right) asymmetry  of distributions of positive and negative  soft pions in the $(k, q_{s}\theta_x)$ plane 
 for the 2011 \MagUp \decay{\Dz}{\Km\pip} subsample, after integration over $\theta_y$.
}
\label{fig:2dim_plot_main}
\end{figure}

A correction to remove the dependence of detection asymmetries on the
soft pion  kinematics is applied in the time-integrated $(k, q_{s} \theta_x,\theta_y)$ distribution, 
where $k=1/\sqrt{p^2_x + p^2_z}$  is proportional to the curvature of the trajectory in the magnetic field,
$q_{s}$ is the sign of the soft pion charge, and $\theta_{x}= \arctan{(p_{x}/p_z)}$, $\theta_y = \arctan{(p_y/p_z)}$ 
are the pion emission angles in the bending and vertical planes, respectively. 
In the absence of any asymmetry in the sample or in the detector acceptance, 
this distribution should be identical for $\Dstarp$ and $\Dstarm$ decays.
A statistically significant asymmetry 
is, however, observed in $\Km\pip$ data (Fig.~\ref{fig:2dim_plot_main}), {where the most visible features are due to geometric boundaries of the detector, where the acceptance for positive and negative tracks differ}.
For each of the three decay modes, 
candidates are therefore weighted to fulfill 
$N^+(k, +\theta_x, \theta_y) = N^-( k, -\theta_x, \theta_y)$, 
where $N^\pm$ is the number of
reconstructed $\Dstarpm$ decays in a given bin. 
The granularity of the correction is
finer in $(k, q_{s}\theta_x)$ than in $\theta_y$, 
where only small non-uniformities are present~\cite{Marino_thesis}.

The weighting procedure corrects for any asymmetry of the detector response, but also removes any global 
asymmetry caused by either \CP violation or differences in the production cross-sections for $\Dstarp$ and $\Dstarm$.
Simulation studies have confirmed that this procedure, while canceling
the time-integrated asymmetry, has no significant effect on a possible genuine time-dependent asymmetry.
The asymmetry correction is independently determined and applied
within each subsample; the convergence of 
all \agamma values for the \Km\pip
control sample to a common value, as seen in 
Fig.~\ref{fig:pseudoAG-raw} \ifthenelse{\boolean{prl}}{(left)}{(top)}, 
thus provides a cross-check of the validity of the method. 
Independent application of the same asymmetry correction procedure to the \decay{\Dz}{\Kp\Km} and \decay{\Dz}{\pip\pim} modes 
also leads to 
good quality for the decay-time fit in each subsample, and good consistency among subsamples, as shown in Fig.~\ref{fig:pseudoAG-raw} \ifthenelse{\boolean{prl}}{(center and right)}{(bottom left and bottom right)}.


Another effect that needs to be accounted for in the measurement of
\agamma\  is the residual contamination from \Dstarp mesons produced
in \bquark-hadron decays. 
This contribution to the measured asymmetry is described with the expression
\begin{equation}
A(t) = (1-\fsec) \aprompt + \fsec \asec, \nonumber
\label{eq:sec_contribution}
\end{equation}
 where \aprompt and \asec are the asymmetries for prompt and secondary  components, and $\fsec$ 
is the fraction of secondary decays in the sample at decay time $t$. 
This fraction is estimated from a simulation-based model calibrated by the yield of secondary decays in data, obtained at high 
values of $t$ from fits to the $\chisqip(\Dz)$ distribution, while \asec is obtained from a 
data sample with $\ln(\chisqip(\Dz))>4$. From these estimates,  the maximum effect of the contamination of secondary decays is assessed as 
$\delta \agamma^{KK} = 0.08 \times 10^{-3}$ and  $\delta
\agamma^{\pi\pi} = 0.12 \times 10^{-3}$,  
accounting for the uncertainty due to the determination of \asec and \fsec, 
and for the possible contribution of non-zero values of $\agamma^{KK}$ and
$\agamma^{\pi\pi}$~\cite{Marino_thesis}.
These effects are much smaller than the statistical uncertainties, 
and are assigned as systematic uncertainties.


Many other effects have been examined as potential sources of systematic uncertainty. 
The uncertainty on the random pion background subtraction has been
evaluated from the measured asymmetry of the background and 
its variation across the mass range surrounding the 
signal peak in the $\Delta m$ distribution,
yielding an uncertainty of $\delta\agamma =0.01 \times 10^{-3}$ for both modes. 
The effect of approximating the continuous, three-dimensional
$(k,q_{s}\theta_x,\theta_y)$ asymmetry correction with a discrete function has been estimated by repeating the extraction of \agamma in the $\Km\pip$ control sample
with twice or half the number of bins, which leads to an uncertainty
of $0.02\times 10^{-3}$ for both decay modes. 
An additional uncertainty in the $K^+K^-$ mode due to the presence of a peaking background 
from real $\Dstarp \to \Dz \pip$ decays with the \Dz meson decaying into other final states
has been evaluated as $\delta \agamma^{KK} = 0.05 \times 10^{-3}$, based on a study of the sidebands of the  \Dz candidate mass distribution.
Other possible sources of systematic uncertainty, including the
resolution of the decay-time measurement, are found to be negligible. 

\begin{figure}[t]
\centering
\ifthenelse{\boolean{prl}}
{
\includegraphics[width=\columnwidth]{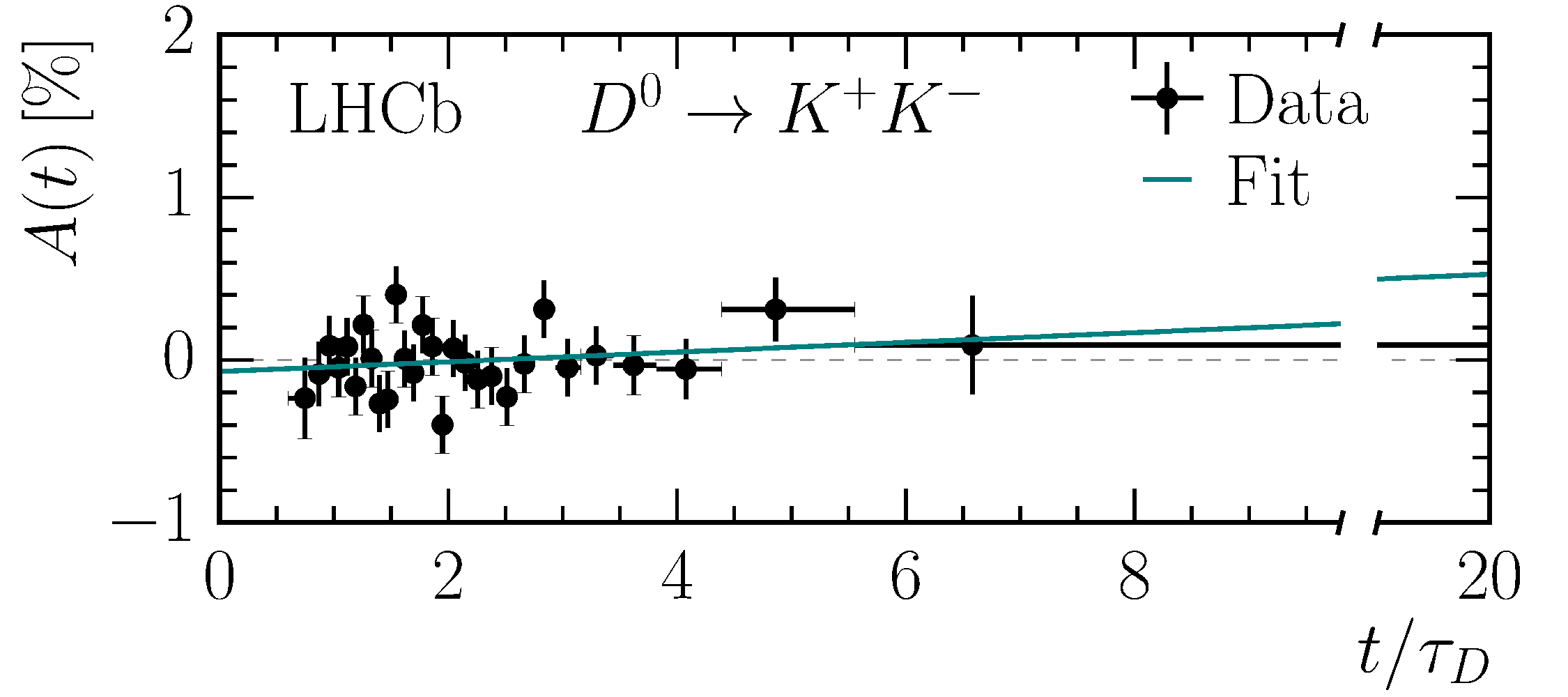}
\includegraphics[width=\columnwidth]{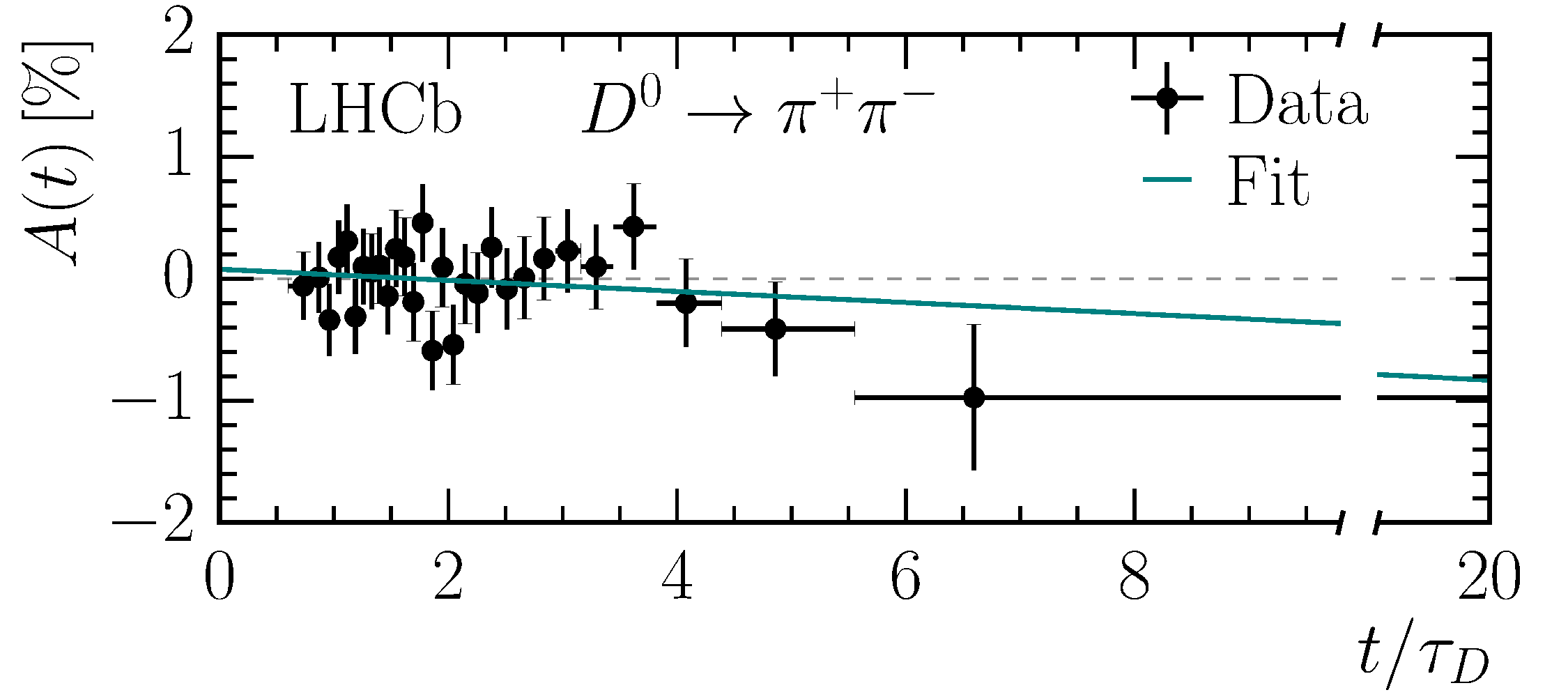}
}
{
\includegraphics[width=0.7\columnwidth]{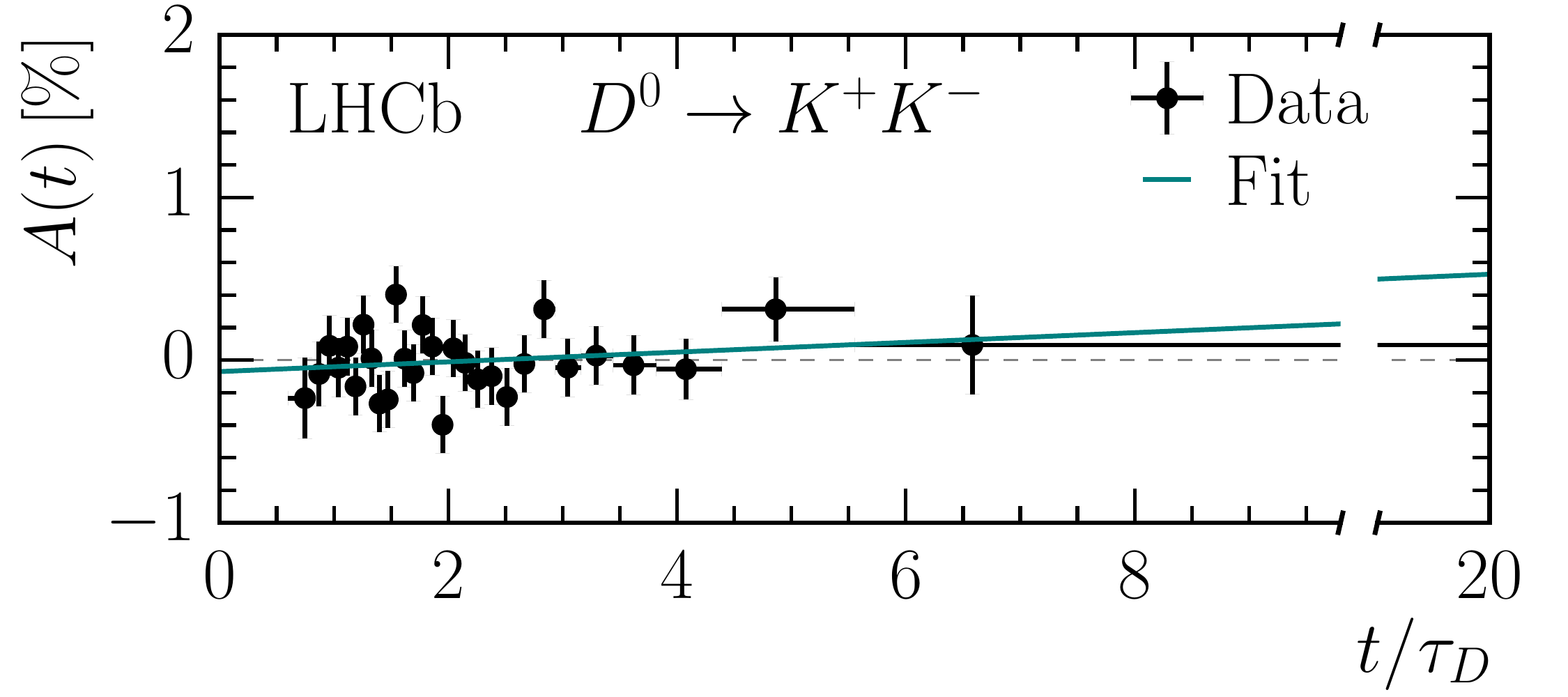}
\includegraphics[width=0.7\columnwidth]{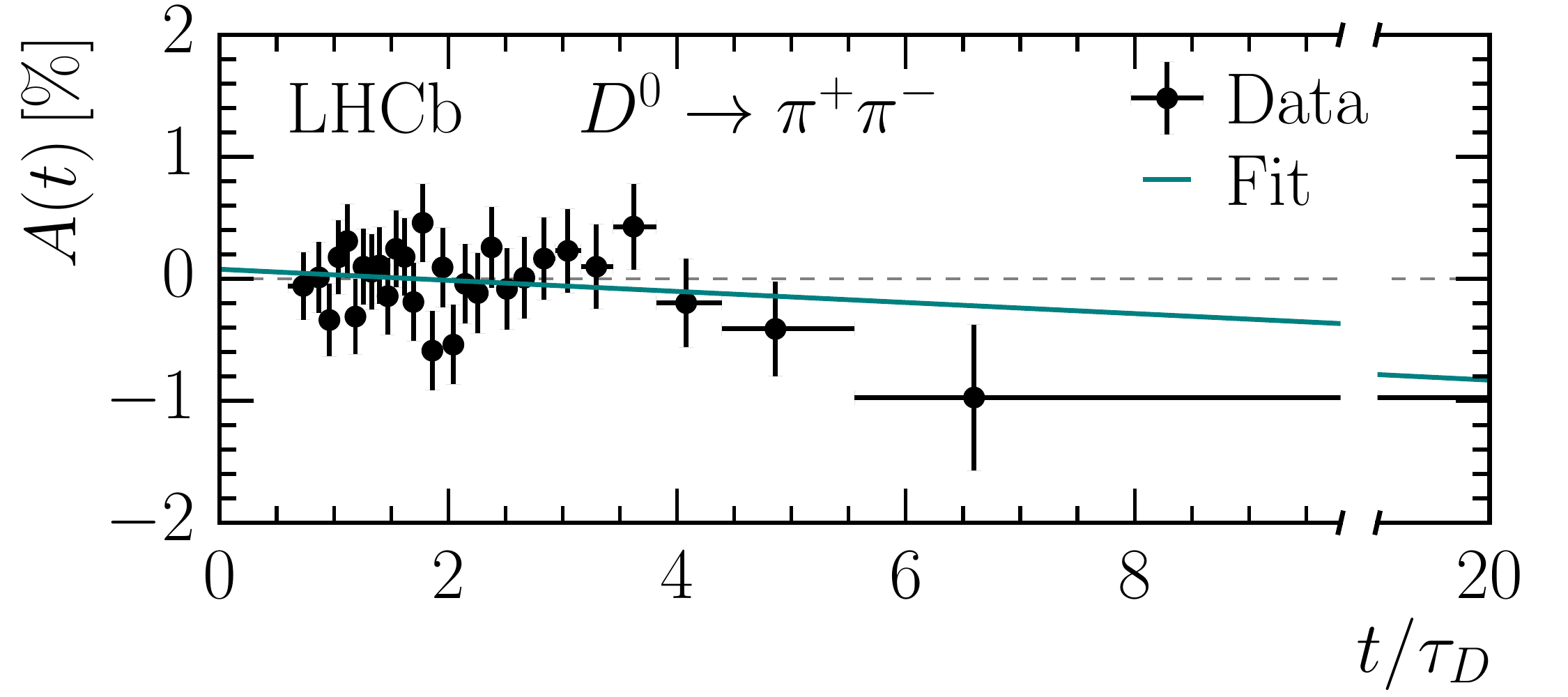}
}
\caption{
Measured asymmetry $A(t)$ in bins of $t/\tau_D$, where $\tau_D = 0.410\ps$~\cite{PDG2016},  
for (top) \decay{\Dz}{\Kp\Km} and (bottom)
  \decay{\Dz}{\pip\pim}, averaged over the full Run~1 data sample.
Solid lines show the time dependence with a slope equal to the best estimates of $-\agamma$.}
\label{fig:results_both}
\end{figure}

The  final results, 
obtained from the weighted average of the values 
separately extracted from time-dependent fits of each subsample (Fig.~\ref{fig:pseudoAG-raw}), are
$\agamma(\Kp\Km) = (-0.30 \pm 0.32 \pm 0.10)\times 10^{-3}$ and 
$\agamma(\pip\pim) = (0.46\pm 0.58 \pm 0.12)\times 10^{-3}$,
where the first uncertainty is statistical and the second is
systematic. Time-dependent asymmetries averaged over the full Run~1
data sample are compared with fit results in Fig.~\ref{fig:results_both}. 

The complementary analysis based on Eq.~\eqref{eq:A_G}
follows a procedure 
largely unchanged from the previous \lhcb
analysis~\cite{LHCb-PAPER-2013-054}, described in
Refs.~\cite{LHCb-PAPER-2011-032,swimming} and  briefly summarized below. 
The selection requirements for this method differ from those based on Eq.~\eqref{eq:Aind}
only in the lack of a requirement on $\chisqip(\Dz)$. 
A similar blinding procedure is used.
This analysis is applied to the $2\invfb$ subsample of the
present data, collected in 2012, that was not used in Ref.~\cite{LHCb-PAPER-2013-054}. The 2012 data is split into three data-taking 
periods to account for known differences in the detector alignment and calibration after detector interventions.

Biases on the decay-time distribution, introduced by the selection
criteria and detection asymmetries, are accounted for through
per-candidate acceptance functions, as described in Ref.~\cite{swimming}. 
These acceptance functions are parametrized by the decay-time intervals within which a candidate would pass the event selection 
if its decay time could be varied. 
They are determined using a data-driven method, 
and used to normalize the per-candidate probability density functions over the decay-time
range in which the candidate would be accepted. 

\begin{figure}[t]
\centering
\ifthenelse{\boolean{prl}}
{
\includegraphics[width=\columnwidth]{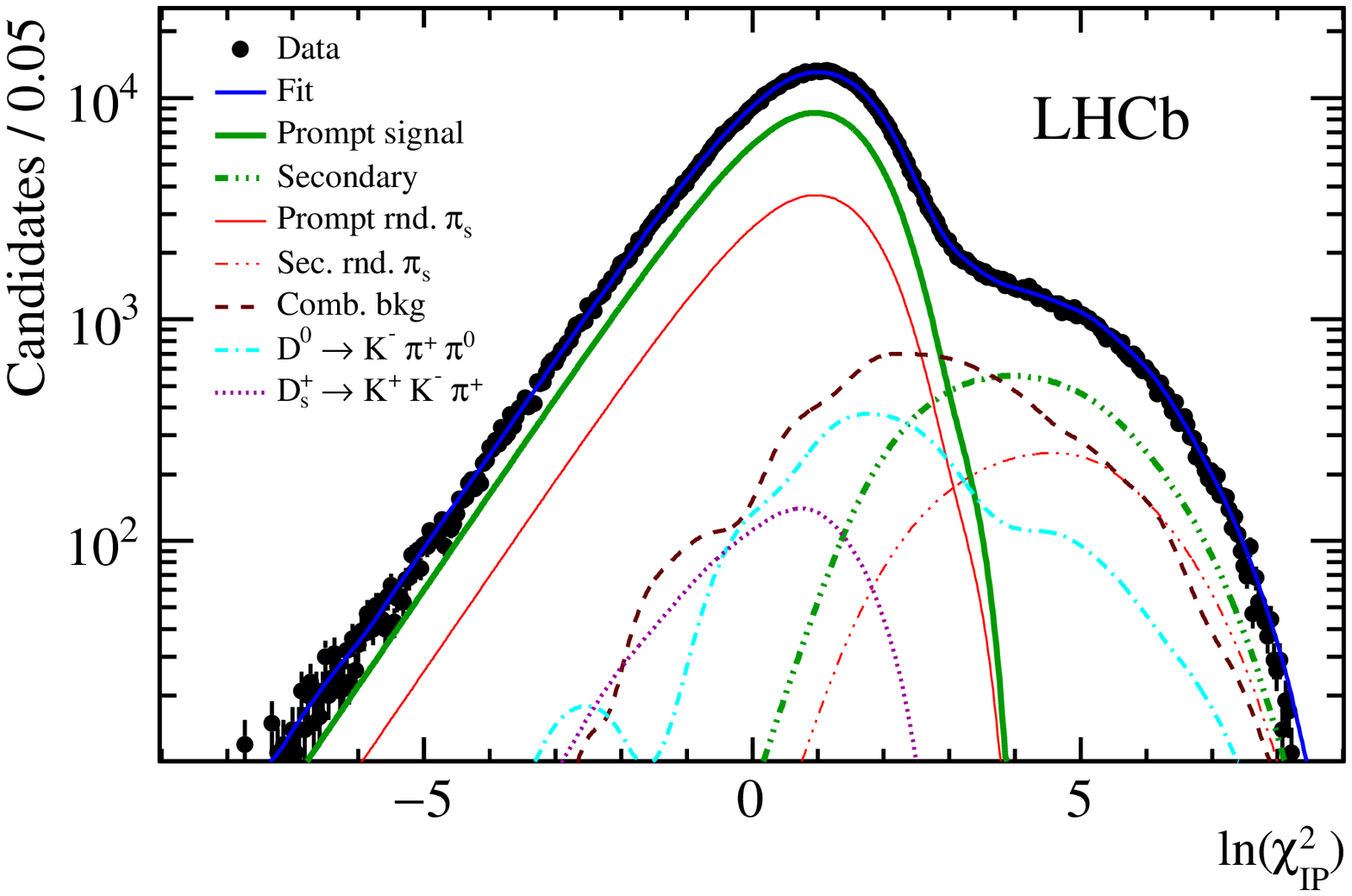}
}
{
\includegraphics[width=0.7\columnwidth]{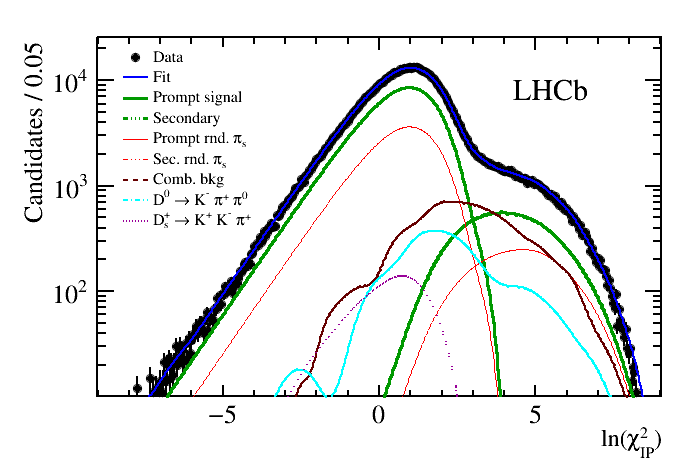}
} 
\caption{Distribution of $\ln(\chisqip(\Dz))$ for the \decay{\Dz}{\Kp\Km} 
candidates selected in the second of the three 2012 data taking periods with magnetic field pointing downwards.
The unbinned maximum likelihood fit results are overlaid. Gaussian kernels are used to smooth the combinatorial and partially reconstructed backgrounds.}
\label{fig:unbinned:lnip_KK_main}
\end{figure}

A two-stage unbinned maximum likelihood fit is used to determine the effective decay widths. 
In the first stage, fits to the \Dz mass and $\Delta m$ spectra are
used to determine yields of signal decays and both combinatorial and partially reconstructed backgrounds. 
In the second stage, a fit to the decay-time distribution together
with $\ln(\chisqip(\Dz))$ (Fig.~\ref{fig:unbinned:lnip_KK_main}) 
is made to separate secondary background.
The finding of an asymmetry consistent with zero in the control channel, $\agamma(\Km\pip)= (-0.07\pm 0.15) \times 10^{-3}$, validates the method.
Small mismodeling effects are observed in the decay-time fits and a corresponding systematic uncertainty
of $0.04\times{10}^{-3}$ ($0.09\times{10}^{-3}$) for \Kp\Km (\pip\pim) is assigned.
The largest systematic uncertainty for the \agamma measurement 
with $\Kp\Km$ ($\pip\pim$)
is $0.08\times{10}^{-3}$ ($0.10\times{10}^{-3}$),  
due to the uncertainty in modeling the contamination from secondary (combinatorial) background.
The results from the 2012 data sample are $\agamma(\Kp\Km,2012) = (-0.03 \pm 0.46 \pm 0.10)\times 10^{-3}$ and $\agamma(\pip\pim,2012) = (0.03\pm 0.79 \pm 0.16)\times 10^{-3}$. These results are then combined with results from Ref.~\cite{LHCb-PAPER-2013-054} to yield the final Run~1 measurements:
$\agamma(\Kp\Km) = (-0.14 \pm 0.37 \pm 0.10)\times 10^{-3}$ 
and $\agamma(\pip\pim) = (0.14\pm 0.63 \pm 0.15)\times 10^{-3}$.

These results can be compared with the  final results from the method based on Eq.~\eqref{eq:Aind}. 
An analysis has been carried out to estimate the statistical
correlation between the results from the two methods, with the conclusion 
that they agree within one standard deviation. 
Due to the large correlation, the measurements from the two methods are not combined, but rather the more precise one is chosen as the nominal result. 

The results for \decay{\Dz}{\Kp\Km} and \decay{\Dz}{\pip\pim} are consistent 
and show no evidence of \CP violation. 
Assuming that only indirect \CP violation contributes to \agamma~\cite{Grossman:2006jg},
and accounting for correlations between the systematic uncertainties~\cite{Nisius:2014wua},   
the two values, obtained with the method using Eq.~\eqref{eq:Aind}, 
can be averaged to yield a single value of
$\agamma = (-0.13 \pm 0.28 \pm 0.10)\times 10^{-3}$, 
while their difference is 
$\Delta \agamma = (-0.76 \pm 0.66 \pm 0.04)\times 10^{-3}$.
The above average is consistent with the result obtained by \lhcb in a muon-tagged sample~\cite{LHCb-PAPER-2014-069}, 
which is statistically independent. 
The two results are therefore combined to yield an 
overall \lhcb Run~1 value 
$\agamma = (-0.29 \pm  0.28)\times 10^{-3}$ 
for the average of the $\Kp\Km$ and  $\pip\pim$ modes.
The measurements of \agamma reported in this Letter are the most
precise to date, and are consistent with previous
results~\cite{LHCb-PAPER-2013-054,Lees:2012qh,Aaltonen:2014efa}. 
They
supersede the previous LHCb measurement~\cite{LHCb-PAPER-2013-054}
with an improvement in precision by nearly a factor of two.

\section*{Acknowledgements}


\noindent We express our gratitude to our colleagues in the CERN
accelerator departments for the excellent performance of the LHC. We
thank the technical and administrative staff at the LHCb
institutes. We acknowledge support from CERN and from the national
agencies: CAPES, CNPq, FAPERJ and FINEP (Brazil); MOST and NSFC (China);
CNRS/IN2P3 (France); BMBF, DFG and MPG (Germany); INFN (Italy); 
FOM and NWO (The Netherlands); MNiSW and NCN (Poland); MEN/IFA (Romania); 
MinES and FASO (Russia); MinECo (Spain); SNSF and SER (Switzerland); 
NASU (Ukraine); STFC (United Kingdom); NSF (USA).
We acknowledge the computing resources that are provided by CERN, IN2P3 (France), KIT and DESY (Germany), INFN (Italy), SURF (The Netherlands), PIC (Spain), GridPP (United Kingdom), RRCKI and Yandex LLC (Russia), CSCS (Switzerland), IFIN-HH (Romania), CBPF (Brazil), PL-GRID (Poland) and OSC (USA). We are indebted to the communities behind the multiple open 
source software packages on which we depend.
Individual groups or members have received support from AvH Foundation (Germany),
EPLANET, Marie Sk\l{}odowska-Curie Actions and ERC (European Union), 
Conseil G\'{e}n\'{e}ral de Haute-Savoie, Labex ENIGMASS and OCEVU, 
R\'{e}gion Auvergne (France), RFBR and Yandex LLC (Russia), GVA, XuntaGal and GENCAT (Spain), Herchel Smith Fund, The Royal Society, Royal Commission for the Exhibition of 1851 and the Leverhulme Trust (United Kingdom).

\addcontentsline{toc}{section}{References}
\setboolean{inbibliography}{true}
\bibliographystyle{LHCb}
\bibliography{main,LHCb-PAPER,LHCb-CONF,LHCb-DP,LHCb-TDR}



 
\newpage
\centerline{\large\bf LHCb collaboration}
\begin{flushleft}
\small
R.~Aaij$^{40}$,
B.~Adeva$^{39}$,
M.~Adinolfi$^{48}$,
Z.~Ajaltouni$^{5}$,
S.~Akar$^{59}$,
J.~Albrecht$^{10}$,
F.~Alessio$^{40}$,
M.~Alexander$^{53}$,
S.~Ali$^{43}$,
G.~Alkhazov$^{31}$
% ...

{\footnotesize \it
% ...
$ ^{5}$Clermont Universit{\'e}, Universit{\'e} Blaise Pascal, CNRS/IN2P3, LPC, Clermont-Ferrand, France\\
% ...
$ ^{10}$Fakult{\"a}t Physik, Technische Universit{\"a}t Dortmund, Dortmund, Germany\\
% ...
$ ^{31}$Petersburg Nuclear Physics Institute (PNPI), Gatchina, Russia\\
% ...
$ ^{39}$Universidad de Santiago de Compostela, Santiago de Compostela, Spain\\
$ ^{40}$European Organization for Nuclear Research (CERN), Geneva, Switzerland\\
% ...
$ ^{43}$Nikhef National Institute for Subatomic Physics, Amsterdam, The Netherlands\\
% ...
$ ^{48}$H.H. Wills Physics Laboratory, University of Bristol, Bristol, United Kingdom\\
% ...
$ ^{53}$School of Physics and Astronomy, University of Glasgow, Glasgow, United Kingdom\\
% ...
$ ^{59}$University of Cincinnati, Cincinnati, OH, United States\\
% ...
}
\end{flushleft}





\end{document}